\documentstyle[11pt,epsfig]{article}
\def \sect #1 {\setcounter{equation} 0\section{#1}}
             
\def \be  {\begin{equation}}
\def \ee  {\end{equation}}
\def \ba  {\begin{eqnarray}}
\def \ea  {\end{eqnarray}}
\def \baa {\begin{eqnarray*}}
\def \eaa {\end{eqnarray*}}
\def \bb  {\begin{thebibliography}}
\def \eb  {\end{thebibliography}}
\newcommand{\bara}{\begin{array}{c}}
\newcommand{\eara}{\end{array}}

\def \lab #1 {\label{#1}}

\def \qqquad {\qquad\quad}
\def \qqqquad {\qquad\qquad}
\def\as{\alpha_S}
\def\np{non-perturbative }
\def\pt{p_{T}}
\def\ptlim{p_{T\,{\rm lim}}} 
\def\blim{b_{\rm lim}}
\def\gev2{{\rm GeV}^2}
\def\nn{\nonumber}

\def\lapproxeq{{\ \lower 0.6ex \hbox{$\buildrel<\over\sim$}\ }}
\def\gapproxeq{{\ \lower 0.6ex \hbox{$\buildrel>\over\sim$}\ }}
\def\pl#1#2#3{
        {\it Phys.\ Lett.\ }{\bf #1} (#2) #3}

\def\prl#1#2#3{
        {\it Phys.\ Rev.\ Lett.\ }{\bf #1} (#2) #3}

\def\pr#1#2#3{
        {\it Phys.\ Rev.\ }{\bf #1} (#2) #3}
\def\nph#1#2#3{
        {\it Nucl.\ Phys.\ }{\bf #1} (#2) #3}
\def\epj#1#2#3{
        {\it Eur.\ Phys.\ J.\ }{\bf #1} (#2) #3}

\def\jhep#1#2#3{
        {\it JHEP }{\bf #1} (#2) #3}

\begin{document}

\vspace*{-2cm}  
\renewcommand{\thefootnote}{\fnsymbol{footnote}}  
\begin{flushright}  
hep-ph/0307208\\
BNL-HET-03/15\\
IPPP/03/40\\
DCPT/03/80\\
July 2003\\  
\end{flushright}  
\vskip 65pt  
\begin{center}  
{\Large {\bf Non-perturbative effects and the resummed Higgs transverse momentum
distribution at the LHC}}\\
\vspace*{1.2cm} 
{\bf  
Anna~Kulesza${}^1$\footnote{Anna.Kulesza@bnl.gov} and  
W.~James~Stirling${}^{2}$\footnote{W.J.Stirling@durham.ac.uk}  
}\\  
\vspace{10pt}  
{\sf 1) Department of Physics, Brookhaven National Laboratory, Upton, NY
11973, U.S.A. \\  
  
2) Institute for Particle Physics Phenomenology, University of Durham,  
Durham DH1 3LE, U.K.}
  
\vspace{70pt}  
\begin{abstract}
We investigate the form of the \np parameterization in both the impact parameter
($b$) space and  transverse momentum ($p_T$) space resummation formalisms 
for the transverse momentum distribution 
of single massive bosons produced at hadron colliders. We propose to analyse
data on $\Upsilon$ hadroproduction as a means of studying the \np contribution in
processes with two gluons in the initial state. We also discuss the theoretical
errors on the resummed Higgs transverse momentum distribution at the LHC arising from the \np contribution. 
\end{abstract}
\end{center}  
\vskip12pt

\setcounter{footnote}{0}  
\renewcommand{\thefootnote}{\arabic{footnote}}  
\vfill  
\clearpage  
\setcounter{page}{1}  
\pagestyle{plain} 


\section{Introduction}

\vspace{5mm} 

The main discovery channel for a light Higgs boson at the future 
Large Hadron Collider (LHC) is the gluon-gluon fusion process $gg \rightarrow HX$.
Studies of the production characteristics are of great importance for further
understanding QCD and determining the Higgs parameters. A
particularly interesting quantity is the transverse momentum ($p_T$) distribution of
the produced Higgs boson. In particular, a precise knowledge of the Higgs $p_T$ distribution is 
important for determining an optimal set of cuts on the final state particles. 
In the past, the study of 
the transverse momentum distribution in  Drell-Yan lepton pair
and electroweak boson production helped to establish the validity of the
soft gluon resummation formalism for the distribution at small $p_T$, and played an
important part in determining the Standard Model (SM) electroweak parameters.

Reliable predictions for transverse momentum distributions at small $p_T$ can only be 
obtained if soft gluon emission effects are correctly taken into account.
The soft radiation manifests itself in the presence of large logarithmic
corrections in the theoretical expressions. The two prominent
examples are recoil logarithms of the form  $\as^n
\ln^{(2n-1)}(Q^2/p_T^2)$ and threshold logarithms of the form of $\as^n
\ln^{(2n-1)} (1-z)/(1-z)$ (where $z=Q^2/\hat{s}$).  A finite result can
be recovered by resumming these corrections to all orders in $\as$.   
To date, the most studied method for resumming the recoil (or `Sudakov') corrections in
transverse momentum distribution for vector boson ($\gamma^*$, $Z$, $W$)
production is the impact parameter ($b$, Fourier conjugate to $p_T$, space)
formalism, also known as the Collins--Soper--Sterman (CSS)
formalism~\cite{CSS}. Derived from the $b$ space resummation formalism are
$p_T$ space methods~\cite{EV,KS,KS2}, which not only provide a very good approximation of
the $b$ space result but also  avoid
certain drawbacks related to the $b$ space method~\cite{EV,KS}. As recently demonstrated, both
the recoil and threshold corrections can be taken into
account within the framework of the joint formalism~\cite{LSV,KSV,KSV2}.

The application of the CSS formalism to Higgs production at the LHC is a straightforward
task and several analyses exist in the literature~\cite{Higgsrecoil, BCS, BQ,
BCDFG}. 
One of the topics discussed here is the application of the $p_T$ space
formalism to Higgs production at the LHC. Some preliminary results based on
the $p_T$ space method have already been presented in Ref.~\cite{LHcontr}.

The CSS formalism resums logarithms arising from arbitrarily soft gluon emission and therefore
needs a prescription for dealing with the Landau pole. Originally this
was incorporated by introducing an arbitrary (inverse energy) scale ($b_*$) 
above which the perturbatively calculated expression remained constant, i.e. `frozen' at
$b_*$. Additionally, a gaussian function, for example $ \exp(-gb^2)$ in its simplest form,
was introduced to correct the
formalism for non-perturbative effects at large $b$, i.e. intrinsic transverse momentum. 
The parameters of this
function are in principle determined from experiment through fits to the data.  
In the case of Drell-Yan lepton pair or electroweak boson
production, determining the size of the \np contribution
associated with the incoming quarks can be
done relatively precisely since there is a significant amount of
experimental data available. 

Other methods for dealing with the Landau singularity have also been proposed
in the literature. The approach of~\cite{LSV,KSV}, initially developed for
the joint
formalism and later applied to the $b$ space formalism in~\cite{BCDFG},
relies on performing the inverse Fourier transform from $b$ to $p_T$ space 
as an integral along a contour in the complex $b$
plane. It returns a well-defined resummed distribution for all nonzero
values of $p_T$, even without introducing any extra \np function. It turns out,
however, that some \np input is still needed to obtain a good description of
the $Z$ production data at the Tevatron~\cite{KSV}. In another development of
the $b$ space formalism, Qiu and Zhang~\cite{QZ} proposed a smooth extrapolation of the
perturbative resummed cross section to push the Landau singularity to infinity.

A complete description of Higgs production within a resummation formalism
also requires  a prescription for dealing with the \np regime. While the
$b_*$ prescription (or other prescriptions) for defining the \np regime 
can be kept the same, a different 
\np function is expected to describe intrinsic $k_T$ in the $gg$ channel. 
A common hypothesis 
is that the \np parameters for this ($gg$) channel can be determined by
multiplying one of the parameters of the \np function for  Drell-Yan ($q \bar q$)
production by a factor of $C_A/C_F$, although to the best of our knowledge there is no rigorous 
theoretical basis for this procedure.

In this paper we calculate the Higgs transverse momentum distribution at the
LHC using the $p_T$ space formalism. 
This method also needs a
non-perturbative (gaussian) input function to account for the intrinsic $k_T$.
In order to determine the non-perturbative coefficients we propose here to
study transverse momentum distribution in $\Upsilon$ production process. We 
argue that the $(gg\to\ \Upsilon)$ hadroproduction provides valuable information on the
amount of intrinsic $k_T$ carried by incoming gluons. The analysis begins
by testing the existing forms of the \np parameterization for Drell-Yan
production and the $C_A/C_F$ hypothesis in $b$ space. A similar study is
then performed for the $p_T$ space formalism.
We finish by considering (or, in the case of $b$ space analysis,
re-examining) the effects of the \np function on Higgs transverse momentum
distribution at the LHC, and draw some conclusions on the uncertainty in the 
prediction for the Higgs boson transverse momentum distribution arising from 
non-perturbative effects.


\section{Theory}

We begin by recalling basic formulae for both $b$ space and
$p_T$ space formalisms. A brief review of \np parameterizations
follows.    


\subsection{$b$ space}
The $b$ space resummed part of the theoretical cross
section for Higgs boson production through the gluon-gluon fusion
in hadronic collisions $p\ p \rightarrow H +X$ has
the following form,
cf.~\cite{CSS}
\begin{eqnarray}
{d \sigma^{\rm res} \over d \pt^2\, d Q^2} &=& {\sigma_0 \tau \pi} \delta(Q^2-m_H^2)  
\int^1_0 \,d x_A \, d x_B \, \delta \left(x_A x_B - {Q^2 \over s}\right)
\times \nn \\
&& {1 \over 2} \int_0^{\infty}  db \, b \,J_0( \pt b) \, \exp[{\cal S}(b_*,Q)]
\,F^{NP}(b,Q,x_A,x_B){f}^{\prime}_{g/A}\left(x_A, {b_0 \over b_*}\right) 
\,{f}^{\prime}_{g/B}\left(x_B, {b_0\over b_*}\right)\,, \nn \\
\label{CSSform}
\end{eqnarray}
with $\sigma_0 = {\sqrt{2} G_F \as^2(\mu) \over 576 \pi}$, $\tau=Q^2/s$, $b_0=2\exp(-\gamma_E)$, and  where
\begin{eqnarray}
{\cal S}(b,Q^2) = - \int_{b_0^2 \over b^2}^{Q^2} \frac{d\bar\mu^2}{\bar\mu^2} 
\bigg[ \ln \bigg ( {Q^2\over\bar \mu^2} \bigg ) A(\alpha_S(\bar\mu^2)) +
B(\alpha_S(\bar\mu^2)) \bigg ] \,,\label{Sbs} \\
A(\alpha_S) = \sum^\infty_{i=1} \left(\frac{\alpha_S}{2 \pi} \right)^i A^{(i)}\
, \quad
B(\alpha_S) = \sum^\infty_{i=1} \left(\frac{\alpha_S}{2 \pi} \right)^i
B^{(i)}\,.
\label{AB}
\end{eqnarray}
For the gluon-gluon fusion process~\cite{KT,DFG},
\ba
A^{(1)} &=&  2 C_A\; , \qqqquad \qqquad
B^{(1)}\;=\;-2 \beta_0\;,\nonumber \\
A^{(2)} &=& 2 C_A \left[
C_A \left( \frac{67}{18}-\frac{\pi^2}{6} \right) -\frac{10}{9}T_R
N_F\right]\; ,
\nonumber\\
B^{(2)}&=& 
C_A^2\left({23 \over 6}+{22 \over 9}\pi^2 -6\zeta_3\right) 
+ 4 C_F N_F T_R - C_A N_F T_R\left({2 \over 3} + {8 \pi^2 \over 9} \right) -
{11 \over 2} C_F C_A \;,
\ea
and $\beta_0= {11 \over 6} C_A -{2\over 3}N_F T_R$.
 
The `modified' parton distributions ${f}^{\prime}$ are 
related to the $\overline{\rm MS}$ parton distributions, $f$,
by a convolution~\cite{CSS,DS,ERV}
\begin{equation}
f^\prime_{g/H} (x,\mu) = \sum_{c}
\int_{x}^{1} {d z \over z} \,
C_{ga}\left( {x \over z},\mu \right)
f_{a/H} \left( z, \mu \right)\ ,
\end{equation}
where
\begin{eqnarray*}
&&\qqqquad C_{ga}(z,\mu) = \sum^\infty_{i=0} \left(\frac{\as}{2 \pi} \right)^i 
C_{ga}^{(i)}\,, \\
C_{gg}^{(0)} (z,\mu) &=& \delta(1-z)\,, \qqqquad \qqqquad \qqqquad \qqqquad C^{(0)}_{gq}(z,\mu)=
0\,, \\
C^{(1)}_{gg}(z,\mu) &=& \delta(1-z)\left[ {\as(\mu) \over 2 \pi} \left( C_A
{\pi^2 \over 6} +  {11\over 2} +\pi^2 
\right) \right]  \,,\qqquad C^{(1)}_{gq}(z,\mu) = {\as(\mu) \over 2 \pi} C_F z \,.
\end{eqnarray*}
The next-to-leading logarithm (NLL) accuracy requires knowledge of the $A^{(1)},\
B^{(1)}, A^{(2)}$ and $C^{(0)}_{ga}$
coefficients; the coefficients $A^{(3)},\ B^{(2)}$ and $C^{(1)}_{ga}$ are of NNLL
order. 

The expression~(\ref{CSSform}) provides a good description of the $p_T$
distribution at small $p_T$; for larger values of $p_T$ one needs to match
the resummed result with the fixed order result. This is achieved by writing
\be 
{d \sigma \over d p_T^2 d Q^2} =  {d \sigma^{\rm res} \over d p_T^2 d Q^2} + Y(p_T,Q) 
\label{match} 
\ee
where $Y(p_T,Q)$ is the difference between the fixed-order and 
resummed results expanded up to the order at which the fixed-order expression
is considered.

The $b_*$ variable in~(\ref{CSSform}) is defined as
\be
b_*= {b \over \sqrt{ 1+ (b/\blim)^2}}\,, \qqqquad b_* <\blim
\label{bstar}
\ee
which ensures that the resummed $b$ space expression is well-defined. The
replacement of $b$ by $b_*$ in~(\ref{CSSform}) prevents the argument of $\as$ from entering the
\np regime by `freezing' the perturbative contribution at a certain
$b \sim \blim$.

Using renormalization group analysis arguments, in~\cite{CSS} a
universal form of the \np function 
$F_{NP}$ was proposed:
\begin{equation}
F^{NP}(Q,b,x_A,x_B) = \exp \left[ -h_Q (b) \ln \left( {Q \over 2 Q_0}
  \right) - h_A (b,x_A) -h_B (b,x_B) \right]\, ,   
\label{CSS:FNP}
\end{equation}
where $Q_0$ is an arbitrary constant indicating the smallest scale at which
perturbation theory is reliable, $Q_0 \sim 1/\blim$. The functions $h_Q,\
h_a,\ h_b$ are to be extracted by comparing theoretical predictions with
experimental data. As postulated in~\cite{CSS}, the flavour dependence of $F^{NP}$ can be
ignored. The $\ln(Q/Q_0)$  dependence in~(\ref{CSS:FNP}) is required to
balance the $Q$ dependence of the Sudakov factor $\exp{\cal S}$. In addition, $h_Q$ was
proved to be universal and its leading $b^2$ behaviour at large $b$ is suggested by the
analysis of the infrared renormalon contribution~\cite{KorS}, as well as a
recent analysis of the dispersive approach to power corrections~\cite{GS}.  

However, the detailed form of the non-perturbative function
$F_{ab}^{NP}(Q,b,x_a,x_b)$ has remained a matter of theoretical dispute. 
Early studies of fixed-target Drell-Yan experimental data, see e.g.~\cite{ESW}, suggested
that a Gaussian parameterization of an intrinsic $q_T$ distribution
provided a good description of data in the low $\pt$ ($1-2$~GeV) regime.
Motivated by this result, Davies et al. (DSW)~\cite{DSW} approximated the
function $F^{NP}$ by
\begin{equation}
F^{NP}(Q,b,x_A,x_B)= \exp \left[ - g_2 b^2 \ln \left( {Q \over 2 Q_0} 
  \right) -  g_1 b^2 \right]\,.
\label{CSS:FNP:DS}
\end{equation}
The $g_1$ parameter in~(\ref{CSS:FNP:DS}) can be interpreted as a measure
of the intrinsic transverse momentum, whereas $g_2$ represents a contribution
coming from unresolved gluons with $k_T<Q_0$ as the structure functions evolve
from scales ${\cal O} (Q_0)$ to ${\cal O} (Q)$.
Assuming~(\ref{CSS:FNP:DS}), with a particular choice of $Q_0=2$ GeV,
$\blim=0.5$ GeV$^{-1}$, and using the Duke-Owens parton distribution
functions~\cite{DO}, the DSW analysis gave
\begin{equation}
g_1=0.15 {\rm\, GeV}^2, \quad \quad  g_2=0.40 {\rm \,GeV}^2\,.
\label{CSS:FNP:DSparam}
\end{equation}

An alternative parameterization, proposed by Ladinsky and Yuan
(LY)~\cite{LY}, incorporates an additional dependence on $\tau= x_a x_b$
\begin{equation}
F^{NP}(Q,b,x_A,x_B)= \exp \left[ - g_2 b^2 \ln \left( {Q \over 2 Q_0} 
  \right) -  g_1 b^2 - g_1 g_3 b\ln (100 x_A x_B) \right]\,.  
\label{CSS:FNP:LY}
\end{equation}
Choosing $Q_0 =1.6$ GeV, $\blim=0.5$  GeV$^{-1}$ and using the CTEQ2M parton
distribution functions, the parameters in (\ref{CSS:FNP:LY}) were determined,
\begin{equation}
g_1=0.11^{+0.04}_{-0.03} {\rm \,GeV}^2, \quad \quad  g_2=0.58^{+0.1}_{-0.2}
{\rm \,GeV}^2, \quad \quad g_3=-1.5^{+0.1}_{-0.1} {\rm \,GeV}^{-1} \,.  
\label{CSS:FNP:LYparam}
\end{equation}

Both parameterizations were reviewed in~\cite{BLLY}. Using
high-statistics samples of Drell-Yan and (CDF Tevatron Run 0) $Z$ production data,
the values of the DSW parameters were updated, 
\begin{equation}
g_1=0.24^{+0.08}_{-0.07} {\rm \,GeV}^2, \quad \quad  g_2=0.34^{+0.07}_{-0.08} {\rm \,GeV}^2\,,
\label{CSS:FNP:DSparamnew}
\end{equation}  
and the LY parameters were found to be
\begin{equation}
g_1=0.15^{+0.04}_{-0.03} {\rm \,GeV}^2, \quad \quad  g_2=0.48^{+0.04}_{-0.05}
{\rm \,GeV}^2, \quad \quad g_3=-0.58^{+0.26}_{-0.20} {\rm \,GeV}^{-1} \,.  
\label{CSS:FNP:LYparamnew}
\end{equation}

Recently, a new form of Gaussian parametrization with an $x$-dependent term,
proportional to $b^2$ (as opposed to the term linear in $b$ in Eq.~\ref{CSS:FNP:LY}) 
has been proposed~\cite{BLNY}:
\be
F^{NP}(Q,b,x_A,x_B)= \exp \left[- g_2 b^2 \ln \left(Q\over 2 Q_0 \right) - g_1 b^2 +
g_1 g_3 b^2 \ln (100 x_A x_B))\right]
\label{BLNY}
\ee
to provide a good description of the low $Q$ Drell-Yan data
 and the CDF, D0 $Z$ data from Tevatron Run 0 and Run I. The values of the
coefficients determined in~\cite{BLNY} are: 
\be
g_1=0.21 \pm 0.01 \ \gev2, \qquad \qquad g_2=0.68 \pm 0.02 \ \gev2, \qquad
\qquad g_3= -0.60^{+0.05}_{-0.04}\ \gev2. 
\label{BLNYparam}
\ee

In our analysis described in the following sections we choose to use the 
standard $b_*$ prescription for the CSS formalism. 
An essentially similar analysis of \np effects can be performed for the CSS 
formalism with different
prescriptions presented in~\cite{LSV,KSV,BCDFG} and in~\cite{QZ,BQ}. 


\subsection{$p_T$ space}
The resummed expression in $p_T$ space, corresponding to~(\ref{CSSform}), has
the following form~\cite{KS2}
\begin{eqnarray}
{ d \sigma \over d \pt d Q^2 } &=& \sigma_0 \tau \pi \delta(Q^2 -M_H^2)
\int_0^1 \,d x_A \, d x_B \, \delta \left(x_A x_B - {Q^2 \over s}\right)
\times \nn \\
&& \Bigg\{ 
-{1 \over p_{T\,*}}  {d p_{T\,*} \over d \pt} \Sigma_1(p_{T\,*}, Q)
f^\prime_{g/A}(x_A,p_{T\,*}) \,f^\prime_{g/B}(x_B,p_{T\,*}) \tilde{F}^{NP}
 \nn \\
&& + \Sigma_2(p_{T\,*},Q) {d p_{T\,*} \over d \pt} {d \over d p_{T\,*}} \left[ 
f^\prime_{g/A}(x_A,p_{T\,*}) \,f^\prime_{g/B}(x_B,p_{T\,*}) \right] \tilde{F}^{NP}
\nn \\
&& + \Sigma_2(p_{T\,*},Q) f^\prime_{g/A}(x_A,p_{T\,*})
\,f^\prime_{g/B}(x_B,p_{T\,*} )
{d \over d \pt} \tilde{F}^{NP} \Bigg\} \,.
\label{KS:FNP}
\end{eqnarray} 
where 
\begin{eqnarray}
\Sigma_2(\pt,Q)&=&\int_{0}^{\infty} dx J_1(x)\exp[{\cal S}(x,\pt,Q)]=\exp({\cal S}_{\eta}) \sum_{N=1}^{\infty} \Bigg({-\alpha_S(\mu^2) A^{(1)}\over
  \pi}\Bigg)^{N-1}{1  \over (N-1)!}\nn
\\
&\times& 
\sum_{m=0}^{N-1} { \left( \begin{array}{c} N-1 \\ m \end{array} \right)} \nn
\sum_{k=0}^{N-m-1} { \left( \begin{array}{c} N-m-1 \\ k \end{array} \right)}
\sum_{l=0}^{N-m-k-1} { \left( \begin{array}{c} N-m-k-1 \\ l \end{array}
  \right)} \nn \\
&\times&
\sum_{j=0}^{N-m-k-l-1} { \left( \begin{array}{c} N-m-k-l-1 \\ j \end{array} \right)}
\sum_{i=0}^{N-m-k-l-j-1} { \left( \begin{array}{c} N-m-k-l-j-1 \\ i
    \end{array} \right)} \nn \\
&\times&
c_2^m c_3^k  c_4^l c_5^j c_6^i  c_1^{N-m-k-l-j-i-1} 
\tau_{N+m+2k+3l+4j+5i-1}\,,
\label{int:hadron}
\end{eqnarray} 
and $\Sigma_1 = -\,\pt\; \partial \Sigma_2\, / \partial \pt$.
The factor ${\cal S}_\eta$ and the $c$ and $\tau$ coefficients are listed
in~\cite{KS}.  

We choose to incorporate the low energy effects using the form 
of the $\pt$ space \np function $\tilde{F}^{NP}$ advocated in~\cite{EV},
\begin{equation}
\tilde{F}^{NP} = 1-\exp{[-\tilde{a}\, \pt^2]}\,. 
\label{F:NP:pt}
\end{equation}
The role of this function is to account for the distribution in the very low
$\pt$ region, and here we are assuming that the shape there is approximately
gaussian. However, in order to combine this with the perturbative result, 
the latter needs to be `frozen'  or `switched off' at some critical 
value of $\pt$ where the coupling $\as$ becomes large. 
A similar freezing is required in the $b$ space approach where the coupling is 
effectively $\as(1/b)$. In other words, in a similar fashion to the $b$ space
method,  we require not only  (i) a form 
$\tilde{F}^{NP}$ for the distribution in the
\np region, but also (ii) a prescription for moving smoothly
from the perturbative to the \np region.  
One possibility for the latter is the 
 `freezing' prescription of~\cite{EV}, 
\begin{equation}
p_{T\,*} = \sqrt {\pt^2 + \ptlim^2 \exp{\left[-\frac{\pt^2}{\ptlim^2}\right]}}\,.
\label{ptstar}
\end{equation}
 which has the property
\begin{equation}
p_{T\,*} =  \left\{ \bara  \pt \, ,   \qquad \qquad \qquad  \quad \pt \gg \ptlim \, , \\
                        \ptlim  \, ,     \qquad \qquad\quad  \quad    \pt \ll \ptlim   \, .  \eara \right. 
\end{equation}
It is important to note that there are {\em two} pieces of information contained
in this definition: the value of the limiting value $\ptlim$ and the abruptness of the transition
 to this value. The use of a gaussian function in the definition (\ref{ptstar}), compared to say
a power law function, implies a rapid transition from the perturbative to the
\np region.

As noted earlier, the considerable amount of data on transverse momentum of
colour singlets produced in processes with two initial quarks makes it
possible to
determine the \np function relatively precisely. This is not the case
for the gluon initiated processes. In fact the only data available are on the 
transverse momentum distribution for low $Q$ resonance production. We argue here that
E605 data on $\Upsilon$ hadroproduction~\cite{E605}  can provide useful information on the
parameters of a \np function for 
processes with {\it gluons} in the initial state. At the values of $x_F$ where the cross section
is measured, the $q\bar q$ contribution to $\Upsilon$ production can be
neglected leaving only the $gg$ channel~\cite{MRSups}. 
Moreover, we also argue that final state
interactions do not introduce significant additional effects and therefore
can be neglected. Since the final state interactions are expected to give rise to 
power corrections, their
effects at values of $Q \sim m_\Upsilon$ relevant for $\Upsilon$ production are at the percent
level of the total cross section. Consequently, the basic mechanisms responsible for 
$\Upsilon$ and Higgs production are similar. We will therefore assume that
a good
description of the transverse momentum in  $\Upsilon$ production can be
achieved with the help of the standard formulae for gluon initiated processes
 in $b$ or $p_T$ space, while allowing for an arbitrary overall normalization.  
We also assume that the major contribution to the measured cross section for
$\Upsilon$ production comes
from the 1S resonance with $m_{\Upsilon(1S)}=9.5$~GeV, and we therefore neglect the 
contributions from other resonances in our theoretical predictions.


\section{Non-perturbative function in $b$ space}

We begin our analysis by studying the non-perturbative function for the impact
parameter formalism in $b$ space. 
We assume the standard value of
the $\blim$ parameter, i.e. $\blim=0.5 $ GeV$^{-1}$. The dependence of a form of
the \np function on the $\blim$ parameter and correlations between the \np
parameters and $\blim$ are not investigated here; a discussion of this issue can be
found in~\cite{ERV}.

Given the $\tau$-independent form of the \np function and the relatively narrow
mass ranges of the data sets under consideration, the fitting procedure
can be greatly simplified if one adopts an effective form for the \np
function
\be
F^{NP}= e^{-g b^2}
\label{fnpeff}
\ee
with $g$ an effective parameter, fitted separately for each of the $p_T$
distributions in the different mass bins. In this way one can test the
dependence of the \np function on $Q$, and in particular possible departures from
the proposed logarithmic dependence. Then, from the effective values of the \np
parameters $g$ at different values of $Q$, the values of the \np parameters $g_1$ and
$g_2$ and/or $g_3$ can be deduced.  All
fitting is done using the MINUIT fitting programme~\cite{MINUIT}. The
resummed predictions fitted to data are accurate up to the NNLL level, but without
including an available numerical estimate for the $A^{(3)}$
coefficient~\cite{AV}, the effect of which is known to be very
small. The $Y$ term in~(\ref{match}) is taken
at leading order. Since all the data analysed are at small $p_T$, the accuracy of matching
does not play a significant role here. 
We use MRST2001 parton distribution functions~\cite{MRST2001} in all our
theoretical predictions.

\begin{figure}[!h]
\begin{center}
\epsfig{figure=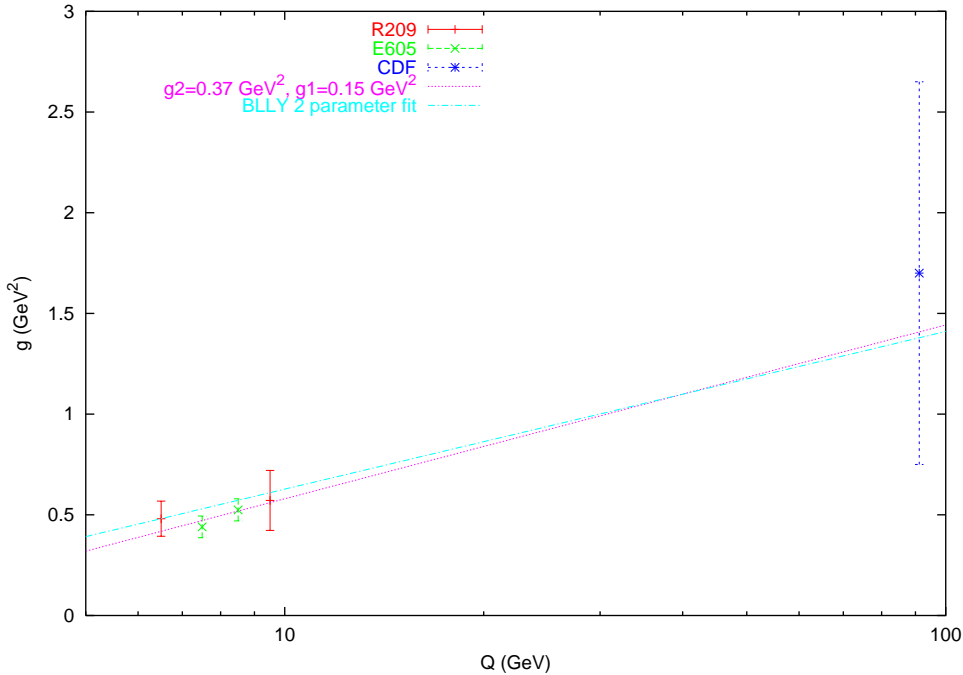,height=13cm,angle=270}
\end{center}
\caption{Two-parameter fit to E605, R209 and CDF $Z$ data (Run 0) data samples
chosen as in~\cite{BLLY}}
\label{run0fig}
\end{figure}

As a test of the fitting method using the effective function~(\ref{fnpeff}), we
attempt to reproduce the values of the $g_1,\ g_2$ parameters obtained by the
BLLY collaboration~\cite{BLLY}. We take exactly the same set of data as in
Ref.~\cite{BLLY}, i.e. the first two mass bins ($7<Q<8$ GeV and $8<Q<9$ GeV)
of the E605 data~\cite{E605} (data with $p_T<1.4$ GeV), R209 data~\cite{R209} ($p_T<2$ GeV) and CDF Run 0 data on $Z$ boson
production~\cite{CDFrun0} (data with $p_T<23$ GeV). The $p_T$ distribution measured by the R209 experiment
in the $8 < Q< 11$ GeV bin was read off the plot in~\cite{BLLY}.
\footnote{ The inclusion of the $8<Q<11$ GeV bin is dubious in any case,
because of contamination by muon pairs coming from $\Upsilon$ decay.} 
The authors of~\cite{ERV} pointed out the impact of the uncertainty in the
normalization of the $p_T$ distributions on the determination of the \np parameters.
Here we deal with the normalization uncertainty by
simultaneously fitting multiple \np effective coefficients $g$ and one common
normalization factor to data gathered in various different mass bins but coming from
the same experiment. Given a set of effective coefficients $g$ for all
mass bins considered, one can investigate its dependence on $Q$. There is evidence
that $g$ increases slowly with $Q$, and in fact
 it is possible to fit a two-parameter function for the effective
parameter $g$ as given by~(\ref{CSS:FNP:DS}), i.e. $g= g_2 \ln \left( {Q /2 Q_0} 
  \right) +  g_1  $, to this data set. 
We obtain $g_1=0.15 \pm 0.13 \ \gev2$, $\ g_2 = 0.37 \pm 0.14\ \gev2$, with $\chi^2/d.o.f.=0.33$. These values are 
within the error range of the BLLY
values~(\ref{CSS:FNP:DSparamnew}) and lead to effective coefficients $g$ for
BLLY and our values of $g_1,\;g_2$ being close to each other over the large range
of $Q$, see Fig.~\ref{run0fig}.

Due to the method of fitting, our errors could be larger then the ones listed
in~\cite{BLLY}. 
The normalization factors for the theoretical predictions with respect to the
 experimental data for the E605 and R209 experiments are
found to be  $0.91 \pm 0.07$ and  $1.06 \pm 0.09$ respectively, and are
in good agreement with~\cite{BLLY}. The normalization of the CDF
$Z$ data from Run 0 is kept equal to one,  as in~\cite{BLLY}. All
our fitting is done using the MINUIT fitting programme~\cite{MINUIT} and
MRST2001 parton distribution functions~\cite{MRST2001}. Given the differences
in the fitting method and the parton distribution functions being used, we interpret our
 ability to recover (within errors) the previously determined
BLLY parameters as an indication that our fitting method can be used as a
valuable diagnostic tool.
The small value of $\chi^2/d.o.f$ may suggest that the
errors on the effective parameters $g$ are overestimated.

\begin{figure}[!h]
\begin{center}
\epsfig{figure=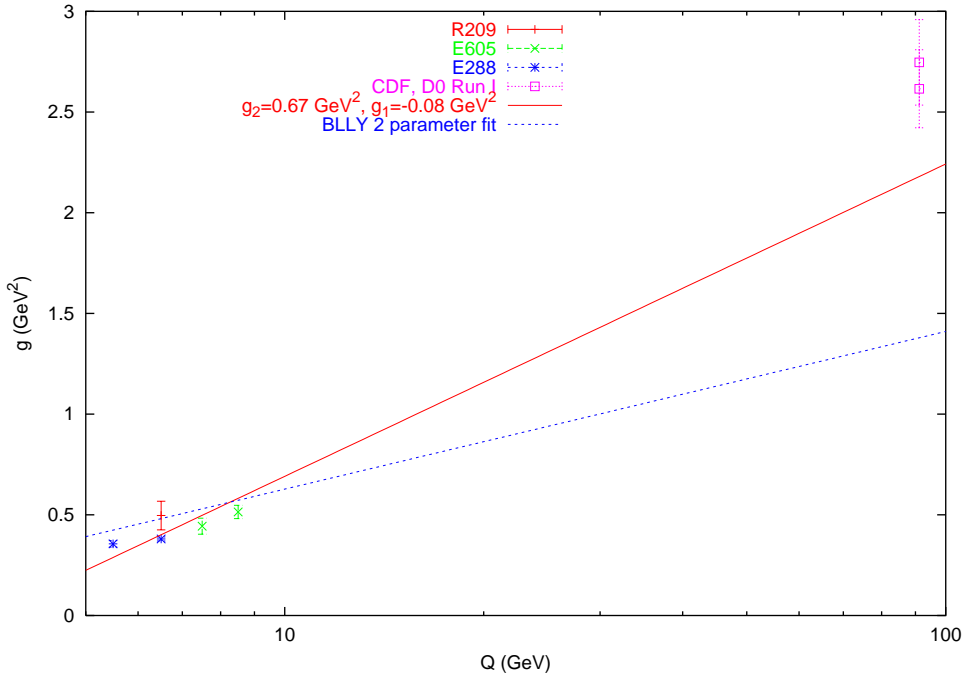,height=13cm,angle=270}
\end{center}
\caption{Two-parameter fit to E288, E605, R209, CDF and D0 $Z$ (Run I) data
samples chosen as described in text.}
\label{run1fig2p}
\end{figure}

However, the BLLY fits were performed using data on $Z$ production gathered
by the CDF collaboration~\cite{CDFrun0} during Run 0 of the Tevatron
$p \bar p$ collider. This particular data set has large statistical errors. The
$p_T$ distribution for $Z$ production was measured much more accurately
during Run I by both the CDF~\cite{CDF:Z} and D0~\cite{D0:Z} collaborations.
A new analysis of the form of the \np function, including the Run I $Z$ data,
appeared recently~\cite{BLNY}. In agreement with the conclusions of~\cite{BLNY}, we
find that the two-parameter form of the $F^{NP}$, given in Eq.~\ref{CSS:FNP:DS}, does
not describe Run I data well, even after refitting for new values of
parameters $g_1$ and $g_2$. In this analysis, in addition to CDF and D0 $Z$
data points with $p_T<20$ GeV, we take R209~\cite{R209}
data points in the $5 \;{\rm GeV} <Q<8 \;{\rm GeV}$ bin only ($p_T<2$ GeV),
E288~\cite{E288} data in the $5 \;{\rm GeV} <Q<6 \;{\rm GeV}$ and $6 \;{\rm GeV} <Q<7
\;{\rm GeV}$ bins ($p_T<2$ GeV, $\sqrt s=27.4$ GeV) and E605~\cite{E605} data in the $7 \;{\rm GeV} <Q<8
\;{\rm GeV}$ and $8 \;{\rm GeV} <Q<9 \;{\rm GeV}$ bins also with $p_T<2$ GeV.
All points from CDF and D0 Run I data sets have
$p_T < 20$ GeV. Due to poor statistics, we do not include the Run 0 data.
Because of known normalization discrepancies between the CDF and D0
experiments, normalization is allowed to be a free parameter while
determining an effective parameter $g$ separately for each data set. 
For data from other experiments, values of effective $g$ parameters are determined in a fit which
allows free normalization factor for theory predictions for each
experiment, but these factors are kept the same for different $Q$ bins from
the same experiment. The normalization factors (multiplicating theoretical results)
are: $N_{E288}=0.81 \pm 0.02,\ N_{E605}=0.90 \pm 0.04,\ N_{R209}=1.1 \pm 0.1,\ N_{CDF}=
1.09 \pm 0.02,\ N_{D0}=0.95 \pm 0.03$. 
The fits of the $b$ space predictions to data return a set of effective $g$
coefficients for all mass bins considered, as shown in
Fig.~\ref{run1fig2p}. A subsequent fit of the function $g= g_2 \ln \left( {Q /2 Q_0} 
  \right) +  g_1  $ to this
set returns $g_1= -0.08\pm 0.09\ \gev2 $, $g_2= 0.67 \pm 0.13\ \gev2$ 
with $\chi^2/d.o.f=8.25$. Fig.~\ref{run1fig2p} illustrates the difference between
an effective coefficient $g$ with these values of
$g_1,\;g_2$ and a coefficient $g$ with the BLLY
values~(\ref{CSS:FNP:DSparam}) as the function of $Q$.

\begin{figure}[!h]
\begin{center}
\epsfig{figure=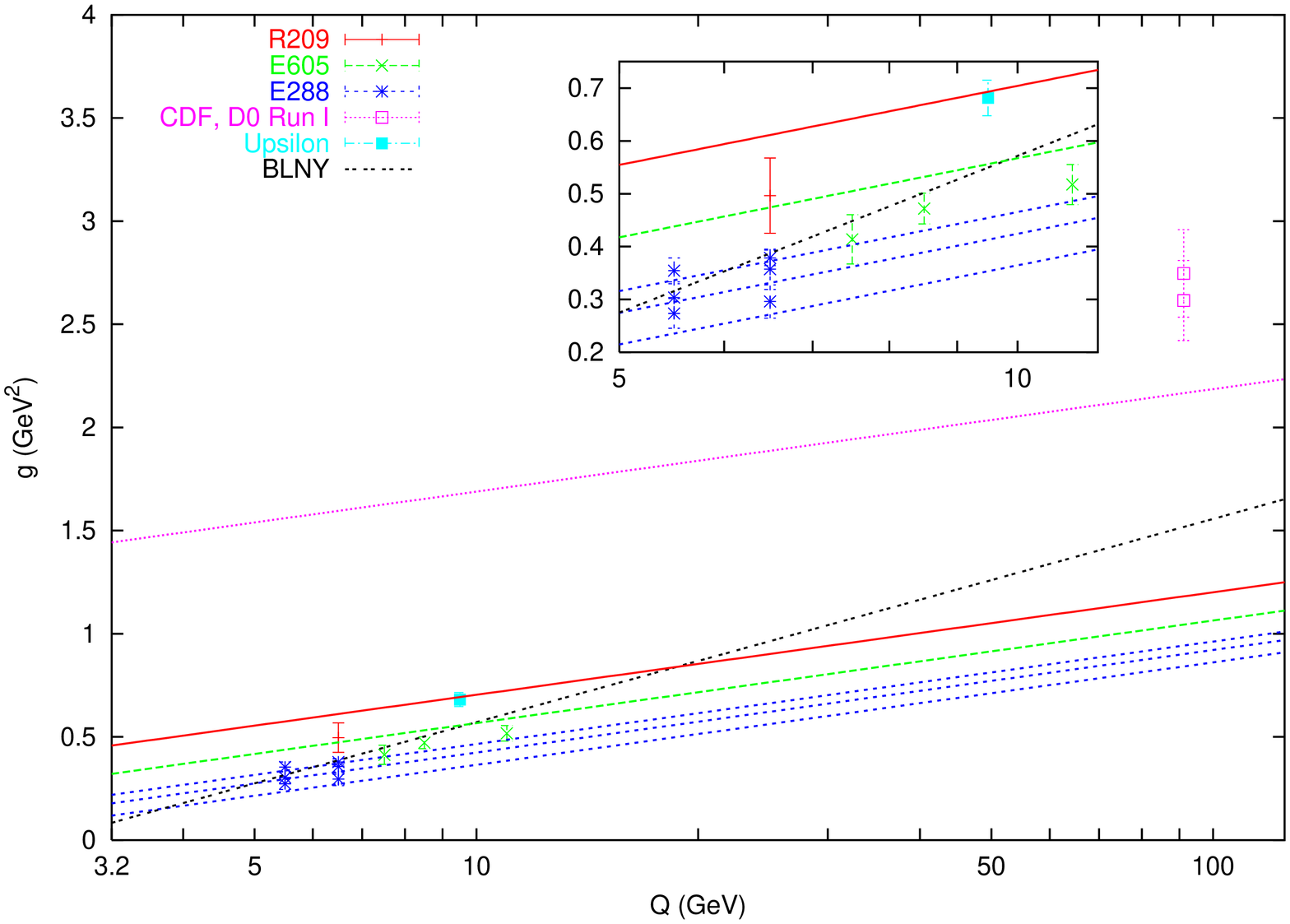,height=13cm,angle=270}
\end{center}
\caption{Three-parameter fit to E288, E605, R209, CDF and D0 $Z$ (Run I) data
samples chosen as described in text. The parallel lines correspond
to \np function of the form~(\ref{rwBLNY}) with
coefficients~(\ref{rwBLNY:param}) and values of $\sqrt s$ of each experiment analysed. The line marked 'BLNY'
corresponds to the BLLY fit of the form~(\ref{BLNY}) with
coefficients~(\ref{BLNYparam}) at $\sqrt s=38.8$ GeV.}
\label{run1fig}
\end{figure}

The BLNY parameterization~(\ref{BLNY}) assumes a gaussian form of the \np
function which includes dependence on $Q$ and $x$. To demonstrate this
dependence in a clear fashion, it is convenient to rewrite the BLNY
parametrization for the effective parameter $g$, cf.~(\ref{BLNY}),
\be
g = g_1 + g_2 \ln \left(Q\over 2 Q_0 \right) + g_1 g_3 \ln \left( 100 x_A x_B
 \right)\;,
\label{geffBLNY}
\ee 
in the form
\be
g = g_1+g_2 \ln \left(Q\over 2 Q_0 \right) + g_3 \ln \left({\sqrt{ s
\over  s_0\;}} \right)\;,
\label{rwBLNY}
\ee 
where we choose $\sqrt s_0 = 19.4$ GeV and $Q_0=1.6$ GeV. In effect, this parameterization 
introduces a $\sqrt s$ dependent component to $g_1$ in the two-parameter form of the 
$F^{NP}$ in Eq.~\ref{CSS:FNP:DS}. 
Apart from the sets of data mentioned
above we also include in the fits testing~(\ref{rwBLNY}) the E288~\cite{E288}
data in the $5 \;{\rm GeV} <Q<6 \;{\rm GeV}$ and $6 \;{\rm GeV} <Q<7
\;{\rm GeV}$ bins ($p_T<2$ GeV) for two additional c.m. energies: $\sqrt s
=19.4,\ 23.8$ GeV, as well as the E605~\cite{E605} data in the $10.5 \;{\rm GeV}
<Q< 11.5$ GeV bin ($p_T<2$ GeV). All the data used in this analysis are
contained in Table~\ref{datatab}. Table~\ref{gvaltab} lists values of
effective $g$ coefficients returned by the fits to the data. They are
also plotted in Fig.~\ref{run1fig}. A subsequent fit
of the function~(\ref{rwBLNY}) to the set of these values reveals
\be
g_1=0.12 \pm 0.07\ \gev2,\qquad \qquad g_2=0.22
\pm 0.12\ \gev2,\qquad \qquad g_3=0.29 \pm 0.09\ \gev2,
\label{rwBLNY:param}
\ee
with $\chi^2 /
d.o.f=3.42$. The best fit normalization factors (multiplicating theoretical results)
are listed in Table~\ref{gvaltab}. Obviously, the quality of the fit improves after introducing 
a third parameter. The quoted value of $\chi^2 /
d.o.f$ also suggests that the choice of logarithmic dependence on $\sqrt s$ in Eq.~\ref{rwBLNY} 
could be too simplified.

Translating the values of the
$g_1,\ g_2$ and $g_3$ parameters above to the BLNY \np function
parameters~(\ref{geffBLNY}) 
leads to 
\be
g_1=0.26 \pm 0.1\ \gev2,\qquad \qquad g_2=0.51
\pm 0.07\ \gev2,\qquad \qquad g_3=-0.55 \pm 0.011\ \gev2,
\label{rwBLNY:param2}
\ee

These values (with the exception of  $g_2$) lie within 
the error band of the parameters determined by the BLNY collaboration. Note,
however, that we use a different fitting technique as well as different data
samples for the analysis.

\begin{table}[!h]
\begin{center}
\begin{tabular}{|c|c|c|c|}           \hline \hline
 Experiment &  $\sqrt s$ (GeV)  & $Q$ range (GeV) & $p_T$ range (GeV)\\ \hline 
  E605      &  38.8             &  7-9, 10.5-11.5 & $0-2$ \\
  E288      &  19.4, 23.8, 27.4 &  5-7            & $0-2$\\
  R209      &  62               &  5-8            & $0-2$ \\
  CDF       &  1800             &  91.2           & $0-20$ \\
  D0        &  1800             &  91.2           & $0-20$ \\ \hline \hline
\end{tabular}
\end{center}
\caption{List of data used in the three-parameter analysis.}
\label{datatab}
\end{table}
\begin{table}[!h]
\begin{center}
\begin{tabular}{|c|c|c|c|c|c|}           \hline \hline
 Experiment & $\sqrt s$ (GeV) &  $Q$ range (GeV) &  effective $g\;(\gev2)$
  & Normalization & $\chi^2/d.o.f$ \\ \hline 
  E605  & 38.8 &  7-8     &  0.414 $\pm$ 0.047 & 0.842 $\pm$ 0.029 & 3.126   \\
        &      &  8-9     &  0.472 $\pm$ 0.029 &                   &         \\
        &      &  10.5-11.5 &  0.517 $\pm$ 0.038 &                 &
  \\ \hline
  E288  & 19.4 &  5-6     &  0.273 $\pm$ 0.028 & 0.807 $\pm$ 0.020 & 1.680 \\
        &      &  6-7     &  0.296 $\pm$ 0.031 &                   &       \\
        & 23.8 &  5-6     &  0.303 $\pm$ 0.029 &                   &       \\
        &      &  6-7     &  0.357 $\pm$ 0.038 &                   &       \\
        & 27.4 &  5-6     &  0.354 $\pm$ 0.024 &                   &       \\
        &      &  6-7     &  0.379 $\pm$ 0.146 &                   &       \\
\hline
  R209  & 62.0 &  5-8     &  0.496 $\pm$ 0.071 & 1.103 $\pm$ 0.096 & 1.515 \\
\hline
  CDF   &1800  &  91.2    &  2.746 $\pm$ 0.212 & 1.090 $\pm$ 0.022 & 0.509
  \\
\hline
  D0    &1800  &  91.2    &  2.615 $\pm$ 0.193 & 0.954 $\pm$ 0.025 & 0.911 \\ \hline \hline
\end{tabular} 
\end{center}
\caption{Effective $g$ coefficients from fits of $b$ space theoretical predictions to data.}
\label{gvaltab}
\end{table}

\begin{figure}[!h]
\begin{center}
\epsfig{figure=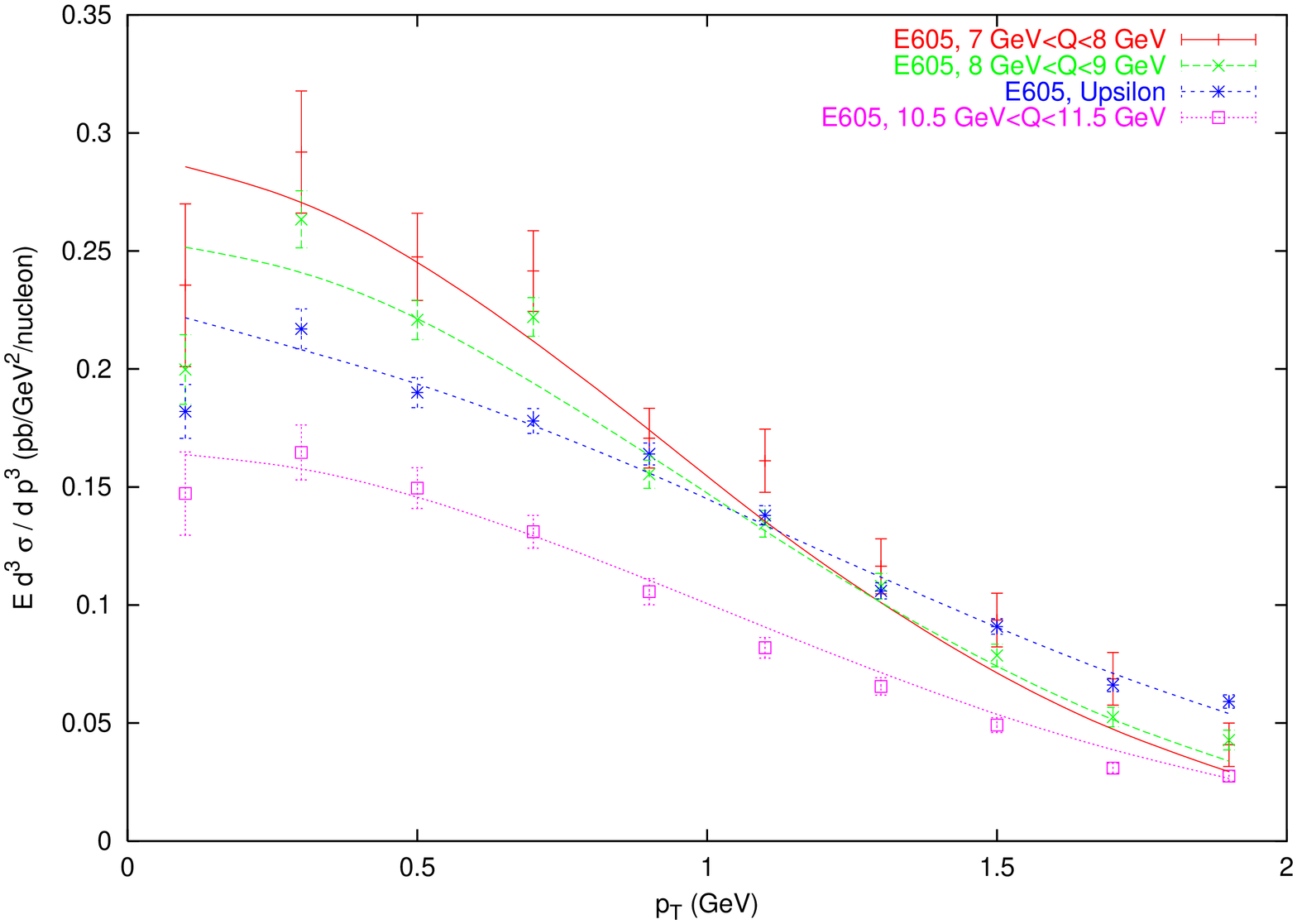,height=13cm,angle=270}
\end{center}
\caption{E605 Drell-Yan (7 GeV$<Q<$8 GeV, 8 GeV$<Q<$9 GeV, 10.5 GeV$<Q<$11.5 GeV bins) and $\Upsilon$ data 
compared to theoretical predictions using the best three-parameter fit.
In order to better compare the shapes
of the $p_T$ distribution, the Drell Yan data have been rescaled by the
factor of 0.3, 0.6, 2.1 for the 7 GeV$<Q<$8 GeV, 8 GeV$<Q<$9 GeV and 10.5
GeV$<Q<$11.5 GeV bins in $Q$, respectively.}
\label{upsidist}
\end{figure}

We next turn our attention to the non-perturbative function for
the case of Higgs production via gluon-gluon fusion, where the transverse
momentum of the Higgs particle at small values of $p_T$ is a result of soft gluon emission
 off the initial-state {\it gluon} lines. In our analysis we use E605 data for $\Upsilon$
 $(pN)$ hadroproduction, allowing the normalization to be a free parameter fitted to data
together with the non-perturbative parameters.
We find the value of the effective parameter $g(\Upsilon)=0.68 \pm 0.03\
 \gev2$ with $\chi^2/ d.o.f =2.79$. Interestingly, this is
relatively close to the values of $g$ for the corresponding $q \bar q$ Drell-Yan production process with
invariant masses of the same order of magnitude, cf. Fig.~\ref{run1fig}.
This conclusion is not unexpected, given the  similar shapes
of the $p_T$ distributions for Drell-Yan and $\Upsilon$ production data, see Fig.~\ref{upsidist}. 

\begin{figure}[!h]
\begin{center}
\epsfig{figure=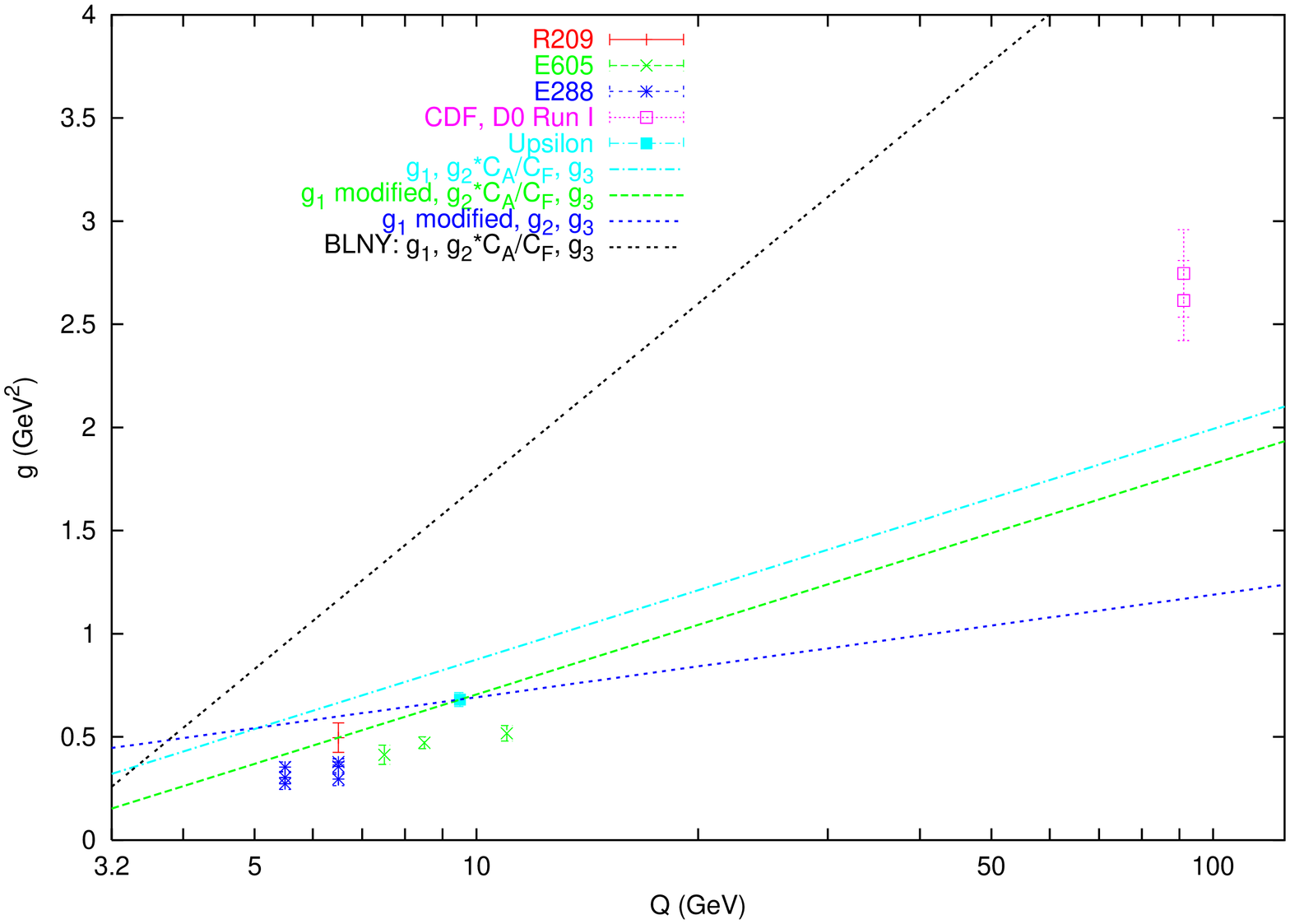,height=13cm,angle=270}
\end{center}
\caption{Drell-Yan (E288, E605, R209), Z production (CDF
and D0 Run 1) and E605 Upsilon data together with the three-parameter fits,~(\ref{BLNY})
and~(\ref{rwBLNY}) at $\sqrt s=38.8$ GeV, with 
coefficients~(\ref{BLNYparam}) and~(\ref{rwBLNY:param}) modified as discussed
in text. }
\label{upsifig}
\end{figure}

The small difference
between the values of $g$ for Upsilon and Drell-Yan production suggests that
the common prescription of multiplying the D-Y coefficient $g_2$ by a factor
$C_A/C_F$ for gluon-initiated processes  
might require adjustments in the
values of the other coefficients in order to maintain the quality of the fit. 
This  is illustrated in
Fig.~\ref{upsifig}, where apart from the effective values of $g$ plotted for
the analysed mass bins, we also plot the fitted effective $g$ at
$\sqrt s = \sqrt s_{\rm E605}=38.8$ GeV,~(\ref{geffBLNY})
and~(\ref{rwBLNY}), with their coefficients $g_2$, as listed
in~(\ref{BLNYparam}) and~(\ref{rwBLNY:param}), multiplied by the factor
$C_A/C_F$. The significant difference between the two predictions originates
not only in the different values of the $g_2$ coefficient obtained in our and
the BLNY fits, but also from the different role of the $g_2$ coefficient itself --
in our case~(\ref{rwBLNY}) it solely measures the dependence on $Q$, whereas
in~(\ref{BLNY}) there is additional $Q$ dependence in the coefficients
$g_1$ and $g_3$ through the relation $x_1 x_2 =Q^2/s$. 
As mentioned earlier, to bridge the gap between experiment and theory some other modifications to
the values of the \np parameters will be needed in addition to the $C_A/C_F$
prescription. In particular, one may argue that while the $g_3$ parameter
should stay the same in the \np functions relevant to processes with $q\bar
q$ and $gg$ initial states, the $g_1$ coefficient can vary depending on the
initial state. The universality of the $g_3$ coefficient can be motivated
by general arguments for the logarithmic dependence on $\sqrt s$ for the contributions
to the $p_T$ of the final state from fluctuations in the underlying event.  
Figure~\ref{upsifig} shows the \np function~(\ref{rwBLNY}) at  $\sqrt
s = \sqrt s_{\rm E605}=38.8$ GeV with $g_1$ modified to obtain
$g=g(\Upsilon)$ both when $g_2$ is  multiplied by $C_A/C_F$ and when it is
left unchanged. The corresponding values of the `adjusted' $g_1$ parameter are
$g_1=-0.05 \pm 0.4\ \gev2$ for the former and $g_1=0.24 \pm 0.23\ \gev2$  for
the latter model. Consequently, the range of allowed values of the effective
parameter $g$ is rather large in each \np model discussed above. Moreover, since 
at the present time it is difficult to discriminate between the various models
experimentally, the values of $g$ from all reasonable models are allowed.  
A band of allowed values of the effective parameters $g$ obtained in this way
can be viewed as a useful estimate of the theoretical error in the $p_T$  distribution
due to the \np input.


\section{Non-perturbative function in $p_T$ space}
  
In general, the analysis of the \np function for the $p_T$
space resummation method follows the  $b$ space analysis described in the previous section, except 
for the additional dependence of the \np function on the parameter $\ptlim$. The value
of this $p_T$ space parameter is {\it a priori}
unknown, and its determination requires simultaneous fitting alongside the other \np
parameters. In $b$ space, the  analogous parameter $b_{\rm lim}$ is
traditionally fixed at a value of 0.5 GeV$^{-1}$. The dependence of the resummed distributions 
for Drell-Yan production on the parameter $b_{\rm lim}$ has been studied in~\cite{ERV}. 
In general, varying the $b_{\rm lim}$ parameter will result in larger errors on the effective 
parameters $g$, and consequently on the transverse momentum distribution itself.   
In the  $p_T$ space case, all the fits we
perform are fits for the effective $\tilde a$ parameters (assumed different for each data
set), the normalization factors (assumed identical for data which come from the same
experiment but have different $Q$) and the factor $\ptlim$ (assumed the same for all data
sets). In fits for the $p_T$ space \np parameterization we use the same data sets and number of data
points as in the $b$ space fits.

\begin{figure}[!h]
\begin{center}
\epsfig{figure=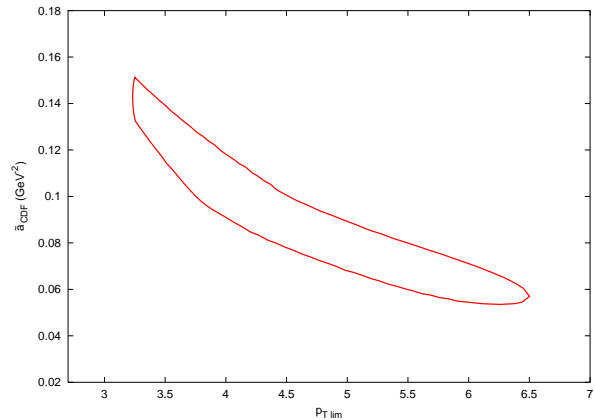,height=8cm,angle=270}
\end{center}
\caption{The 68\% C.L. contour in the $\ptlim$,\ $\tilde a_{\;CDF}$ plane.}
\label{chi2ellipse}
\end{figure}

From a simultaneous fit to low $Q$ Drell-Yan and Run I
$Z$ production data, we establish the value of $\ptlim$ to be
$\ptlim=5.5 \pm 2.98$ GeV with $\chi^2/d.o.f=1.54$.
The value of $\ptlim$ is strongly correlated with the normalization factors
for the low $Q$ experiments and the value of $\tilde{a}$ for $Z$ production
data, see Fig.~\ref{chi2ellipse}. The normalization factors listed in
Table~\ref{avaltab} are
different from their $b$ space counterparts. The effective parameters
$\tilde{a}$ for all $Q$ bins are now determined in one fit 
which also determines the common factor $\ptlim$. In the $b$ space analysis, the
effective parameters $g$ for each $Q$ bin resulted from separate fits for
all experiments -- we did not fit for a common factor $b_{\rm lim}$.
The resulting value of $\ptlim$ appears rather large, if interpreted 
as defining the border between perturbative and \np physics.

Motivated by the form of the three-parameter \np function~(\ref{rwBLNY}) in $b$ space,
 we propose the following form for the dependence of the effective \np parameter $\tilde{a}$
on $Q$ in~(\ref{F:NP:pt}):
\be
\tilde{a}=\left( \tilde{a}_1 + \tilde{a}_2 \ln \left(\frac{Q}{2Q_0}\right)+
\tilde{a}_3 \ln \left( \sqrt{\frac{s}{s_0}}\right)\right)^{-1}
\label{afnp}
\ee

\begin{figure}[!h]
\begin{center}
\epsfig{figure=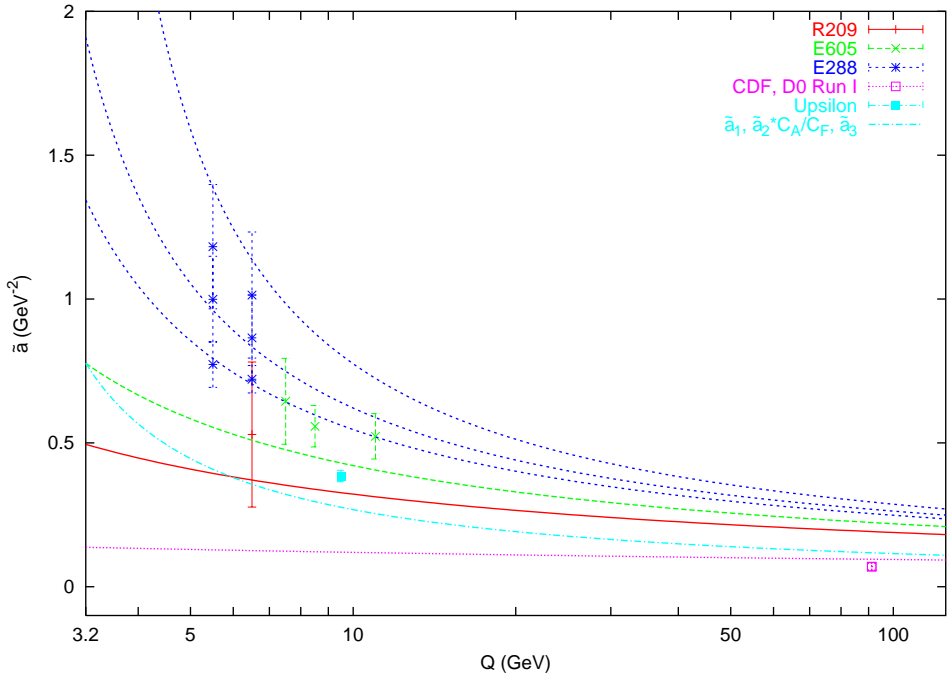,height=13cm,angle=270}
\end{center}
\caption{Effective parameters $\tilde{a}$ for the Drell-Yan (E288, E605, R209), Z
production (CDF and D0 Run 1) and E605 Upsilon data and the \np
parameterization~(\ref{afnp}), together  with the best fit values of
coefficients~(\ref{afnp:param}) plotted for the $\sqrt s$ of each set 
of experimental data analysed.}
\label{aval}
\end{figure}

As for the $b$ space analysis, we first fit the theoretical predictions to
data in order to determine the effective coefficients $\tilde a$ for all mass
bins analysed.  
A fit of (\ref{afnp}) to the set of effective values of $\tilde{a}$,  listed in 
Table~\ref{avaltab} and  
shown in Fig.~\ref{aval}, returns the following values for the  parameters:
\be 
\tilde{a}_1=0.20 \pm 0.50\; {\rm GeV}^{-2},\qquad \qquad \tilde a_2=0.95\pm 0.92
\; {\rm GeV}^{-2}, \qquad \qquad\tilde{a}_3=1.56 \pm 0.57\; {\rm GeV}^{-2},
\label{afnp:param}
\ee
with $\chi^2/d.o.f=1.38$.
The errors on the coefficients are substantially larger than the corresponding 
errors on the coefficients in $b$ space, again due to the much larger number of
parameters (i.e., the effective values of $\tilde{a}$ for bins in $Q$) determined in
a single fit in the $p_T$ space analysis as compared to the $b$ space fits. In
Fig.~\ref{aval} we also plot the \np parameterization~(\ref{afnp}) for the $\sqrt
s$ of each experiment analysed here, and the best fit coefficients
$\tilde{a}_1, \tilde{a}_2, \tilde{a}_3$ listed above. As for $b$ space,
the three parameter fit provides a much better description of the data than the
two parameter case.
\begin{table}[!h]
\begin{center}
\begin{tabular}{|c|c|c|c|c|c|}           \hline \hline
 Experiment & $\sqrt s$ (GeV) &  $Q$ range (GeV) &  effective $\tilde{a}
  \;({\rm GeV}^{-2})$ & Normalization & $\chi^2/d.o.f$ \\ \hline 
  E605  & 38.8 &  7-8     &  0.645 $\pm$ 0.149 & 0.928 $\pm$ 0.067 & 1.552  \\
        &      &  8-9     &  0.558 $\pm$ 0.072 &                   &         \\
        &      &  10.5-11.5 &  0.523 $\pm$ 0.079 &                 &
  \\ \cline{1-5}
  E288  & 19.4 &  5-6     &  1.182 $\pm$ 0.216 & 0.991 $\pm$ 0.050 &       \\
        &      &  6-7     &  1.014 $\pm$ 0.219 &                   &       \\ \cline{2-5}
        & 23.8 &  5-6     &  0.998 $\pm$ 0.149 &                   &       \\
        &      &  6-7     &  0.865 $\pm$ 0.149 &                   &       \\\cline{2-5}
        & 27.4 &  5-6     &  0.773 $\pm$ 0.080 &                   &       \\
        &      &  6-7     &  0.721 $\pm$ 0.047 &                   &       \\ \cline{1-5}
  R209  & 62.0 &  5-8     &  0.529 $\pm$ 0.252 & 1.141 $\pm$ 0.288 &     \\ \cline{1-5}
  CDF   &1800  &  91.2    &  0.070 $\pm$ 0.012 & 1.050 $\pm$ 0.065 & 
  \\ \cline{1-5}
  D0    &1800  &  91.2    &  0.070 $\pm$ 0.013 & 0.914 $\pm$ 0.072 &  \\ \hline \hline
\end{tabular} 
\end{center}
\caption{Effective coefficients $\tilde{a}$ from fits of the $p_T$ space
theoretical predictions to data.}
\label{avaltab}
\end{table}

Given the $b$ space results for the fit to the $\Upsilon$ data, we would  expect the
effective parameter $\tilde{a}$ for $\Upsilon$ production to be relatively close
to the corresponding values of $\tilde{a}$ for Drell-Yan production. This is
indeed the case, as can be seen from Fig.~\ref{aval}. The fitted
$\tilde{a}$ value for Upsilon production is $\tilde{a}(\Upsilon) =0.39 \pm 0.02$
GeV$^{-2}$ ($\chi^2/d.o.f.=2.70$). The value of $\tilde{a}(\Upsilon)$ is
determined from a simple fit to the $p_T$ distribution data
where the only other quantity fitted is the overall normalization;
$\ptlim$ is kept fixed at the value determined from the fits to the DY data and
the error on $\ptlim$ is not taken into account.
The resulting error on the fitted value of $\tilde{a}(\Upsilon)$ is therefore 
superficially small. Nevertheless, it remains true that the central value of
the effective $\tilde{a}(\Upsilon)$ is closer to the
value predicted by the \np parametrization~(\ref{afnp})
with~(\ref{afnp:param}) than by the same parameterization with the
$\tilde{a}_2$ coefficient multiplied by $C_A/C_F$ -- the equivalent of the $b$
space prescription of rescaling the $g_2$ coefficient by $C_A/C_F$. It therefore
becomes an attractive alternative to consider other possible modifications to
the values of the \np parameters, determined for the Drell-Yan type processes, to
describe processes with gluons in the initial state. Following the $b$ space
analysis presented in the previous section, we make the choice of adjusting the
$\tilde{a}_1$ parameter while keeping  $\tilde{a}_3$ unchanged and the parameter
$\tilde{a}_2$ either unchanged or rescaled by $C_A/C_F$. The resulting values of
the parameter  $\tilde{a}_1$ are:  $\tilde{a}_1=0.47^{+1.53}_{-1.52}$
GeV$^{-2}$ and  $\tilde{a}_1=-0.82^{+2.78}_{-2.77}$ for the former and
latter model, respectively. The magnitude of the errors is a straightforward
consequence of the large errors on the effective parameters $\tilde {a}$ for
the bins in $Q$.


\section{The Higgs transverse momentum distribution in the $p_T$ space formalism}

As discussed previously, in the resummation framework the $p_T$ distribution for the gluon-fusion 
induced Higgs production
process is obtained assuming that the \np contribution has the same form
as the \np function for
Drell-Yan type processes --- but with the $g_2$ or a corresponding
parameter rescaled by a factor of $C_A/C_F$. 
The results presented in the previous two
sections may suggest exercising some caution when applying this hypothesis.

In this section we focus on the effect of the \np function on predictions 
 for the Higgs $p_T$ distribution, in both the $b$ space and $p_T$ space
 approach. In addition, we discuss the changes made to the coefficients of the
 \np function(s) to allow for 
 gluons in the initial state and the impact of these modifications on
the predictions for the Higgs $p_T$ distribution.

\begin{figure}[!h]
\begin{center}
\epsfig{figure=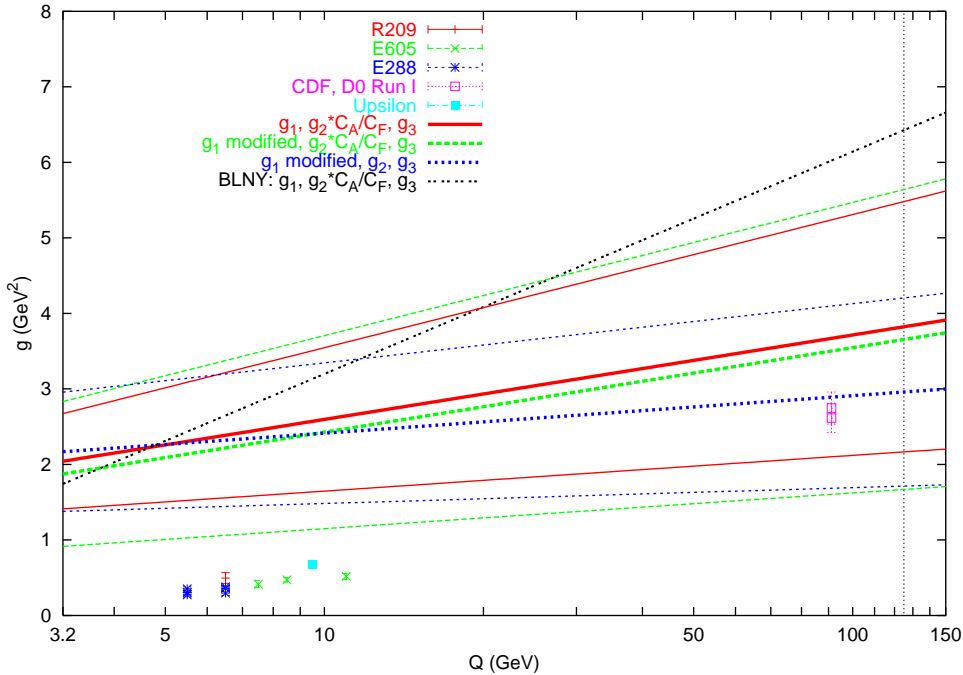,height=13cm,angle=270}
\end{center}
\caption{The effective parameter $g$ as a function of $Q$ in various
models of the \np parametrization  in $b$ space at $\sqrt s =14$~TeV. The central values
are denoted by thick lines; the errors bands by thin lines. The dotted
vertical line is drawn for $Q=125$~GeV.}
\label{gvalhiggs}
\end{figure}

First, we focus on the $b$ space \np function. As can be seen from
Fig.~\ref{gvalhiggs}, the central values of the effective parameters $g$ for
$\sqrt s =14$~TeV   are rather close to each other, irrespective of the different treatments 
of the $g_1,\;g_2,\;g_3$ coefficients discussed 
above. However, the significant errors on the coefficients 
$g_1$, $g_2$ and $g_3$ result in an even more pronounced error on the effective
parameter $g$ in each model. The errors are larger in models in which the
$g_2$ coefficient is rescaled by $C_A/C_F$, as the errors get rescaled together with the
central values. If we assume the mass of the Higgs boson to be $M_H=125$ GeV, the 
effective values are: $g=3.82 \pm 1.66 \ \gev2$ for the model with $g_2$
rescaled by $C_A/C_F$ and $g_1$, $g_3$ unchanged, $g=3.65 \pm 1.99 \ \gev2$ for
the model with $g_2$ rescaled by $C_A/C_F$, $g_1$ modified and $g_3$ unchanged,
$g=2.96 \pm 1.25 \ \gev2$ for the model with $g_2$, $g_3$ unchanged and $g_1$
modified. 
The next step is to implement the \np function~(\ref{fnpeff}) with
these effective values of $g$ into the resummed formalism. Again, we view the
spread in the central values of $g$ due to various models, together with the  
error bands on $g$ in each model, as a measure of the theoretical error
on the effective parameter $g$. Consequently, we present the Higgs $p_T$ distributions
for the smallest and the largest value of 
$g$ allowed by the error bands for all models. These values both happen to
come from the model with $g_2$ rescaled by $C_A/C_F$, $g_1$ modified
and $g_3$ unchanged, and correspond to  $g=1.67\ \gev2$ and $g=5.64\ \gev2$.
 
\begin{figure}[!h]
\begin{center}
\epsfig{figure=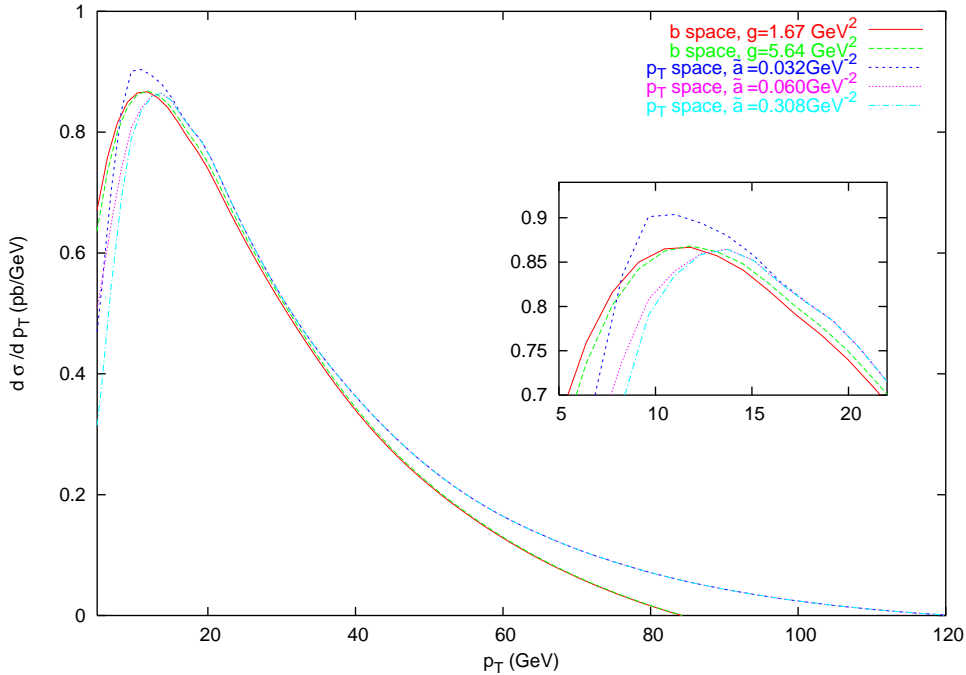,height=13cm,angle=270}
\caption{Higgs boson $p_T$ distributions at the LHC, as predicted by the $b$
space and $p_T$ space resummation formalism.}
\label{higgs}
\end{center}
\end{figure}

Figure~\ref{higgs} shows the predicted transverse momentum distribution of the Higgs
boson at the LHC calculated in the framework of the standard $b$ space
resummation with the effective \np parametrization of the form~(\ref{fnpeff})
for two `extreme' values of $g$ listed above. 
The resummed predictions are accurate up to NNLL level, i.e. they include the
NNLL coefficients $C^{(1)},\;B^{(2)}$ as well as a numerical estimate of the
coefficient $A^{(3)}$~\cite{AV}, rescaled by $C_A/C_F$. The matching is
performed at the leading-order level.
We assume $M_H=125$~GeV and
use MRST2001 parton distribution functions.   
We find, in agreement with Ref.~\cite{BCS,BQ}, only a very small
dependence of the theoretical predictions on the exact values of the \np
parameters. The predictions shown in Fig.~\ref{higgs} are somewhat lower than
the ones presented in~\cite{BCDFG} but higher then those obtained
in~\cite{BQ}. 

\begin{figure}[!h]
\begin{center}
\epsfig{figure=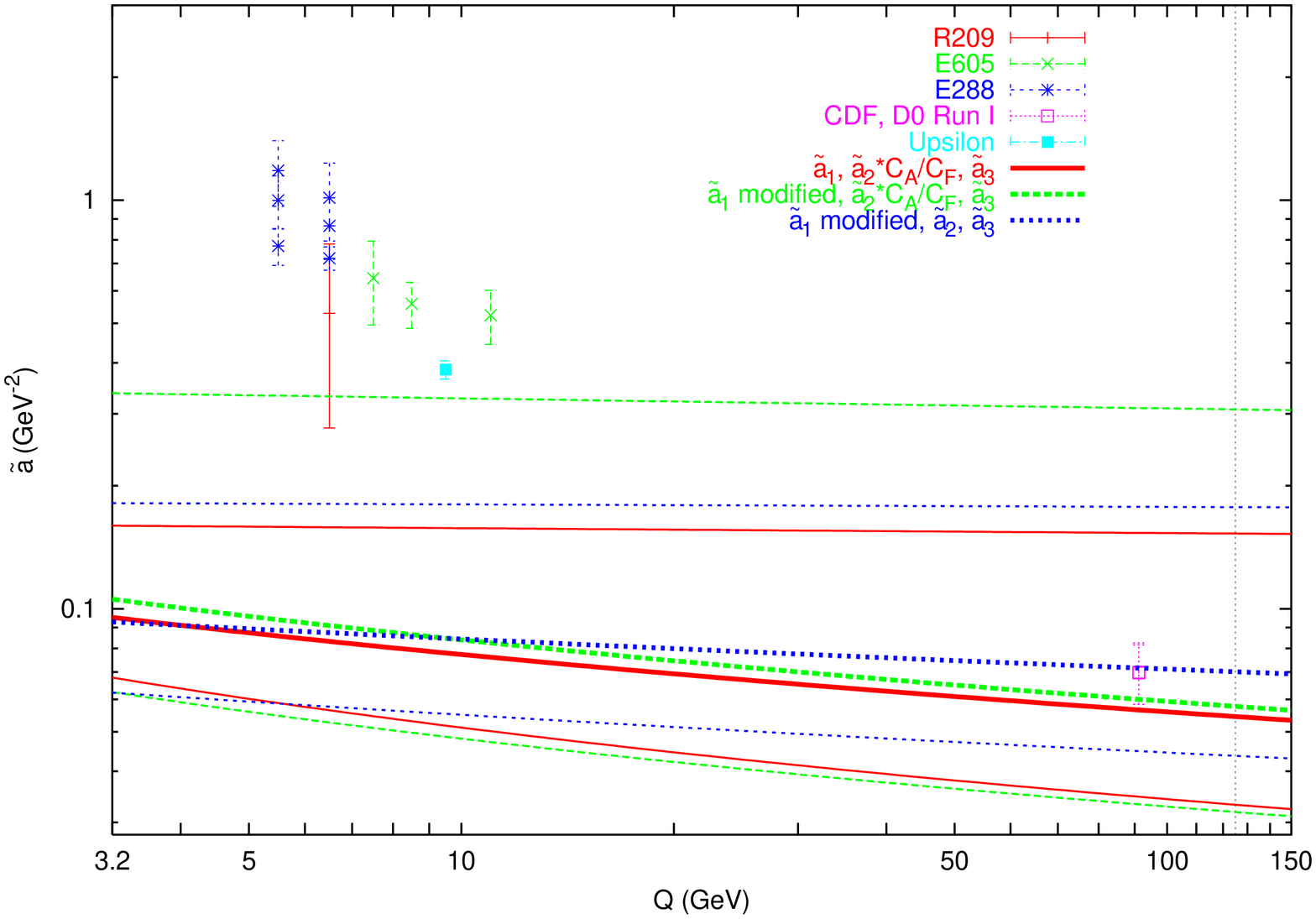,height=13cm,angle=270}
\end{center}
\caption{The effective parameter $\tilde{a} $ as a function of $Q$ in various models of
the \np parametrization  in $p_T$ space at $\sqrt s=14$~TeV. The central values
are denoted by thick lines; the errors bands by thin lines. The dotted
vertical line is drawn for $Q=125$~GeV.}
\label{avalhiggs}
\end{figure} 

Analogously to $b$ space, the central values of the $p_T$-space parameter $\tilde a$, as predicted by
the different models for $\sqrt s=14$~TeV and $M_H=125$~GeV, lie relatively close
to each other, see Fig.~\ref{avalhiggs}. However, due to the large errors on the
central values of $\tilde a$ associated with each model, the range of
possible values of $\tilde a$ is an order of magnitude large. The central
values (for $M_H=125$~GeV) are: $\tilde a=0.054^{+0.098}_{-0.021}$ GeV$^{-2}$ for the model with $\tilde{a}_2$
rescaled by $C_A/C_F$ and $\tilde{a}_1$, $\tilde{a}_3$ unchanged,
$\tilde{a}=0.058^{+0.250}_{-0.026}$ GeV$^{-2}$ for
the model with $\tilde{a}_2$ rescaled by $C_A/C_F$, $\tilde{a}_1$ modified and $\tilde{a}_3$ unchanged,
$\tilde{a}=0.070^{+0.107}_{-0.026}$ GeV$^{-2}$ for the model with $\tilde{a}_2$,
$\tilde{a}_3$ unchanged and $\tilde{a}_1$ modified. The smallest and largest
values of $\tilde a$  also happen to come from the model with $\tilde a_2$
rescaled by $C_A/C_F$, $\tilde a_1$ modified
and $\tilde a_3$ unchanged, and are: $\tilde{a}=0.032$ GeV$^{-2}$ and
$\tilde{a}=0.308$  GeV$^{-2}$.

The Higgs transverse momentum distribution obtained in the framework of the
$p_T$ space formalism is compared to the $b$ space results in
Fig.~\ref{higgs}. The $p_T$ space distribution is calculated assuming the effective \np
function of the form~(\ref{F:NP:pt}) with a coefficient $\tilde a=0.06$
GeV$^{-2}$ and the best fit value of $\ptlim=5.5$ GeV. The value of $\tilde
a=0.06$ GeV$^{-2}$ is an average of the central values of $\tilde a$ in
the various models of treating the $\tilde a_1,\;\tilde a_2,\;\tilde a_3$ coefficients discussed above. The $p_T$ space prediction agrees
reasonably well with the $b$ space prediction in the small $p_T$ regime.
The gap between the locations of the peaks of the distributions gets narrower for
smaller values of $\ptlim$ as well as smaller values of $\tilde a$. At larger values of $p_T$, the predictions of
the  two
formalisms differ substantially. It is well known that the $b$ space
distribution, even after matching with the fixed-order result, is eventually 
going to turn negative at sufficiently large $p_T$. This happens because 
the matching relies on the perturbative fixed-order result, known up to a certain
order in $\as$. At one order higher than the matching accuracy, the resummed
expression introduces logarithmic terms which will not be cancelled by the
terms coming from the matching term $Y$ in~(\ref{match}). 
Since here matching is performed only
at the LO level, the negative behaviour of the matched resummed 
distribution at large $p_T$ is quite pronounced. In the numerical realisation of the $p_T$ space method, only a
subset of subleading higher order terms is summed. Consequently, the matched resummed
distribution becomes negative at much larger values of $p_T$, increasing the
range of applicability of the resummation approach. 

As the analytic results for
the NLO $p_T$ Higgs distribution are available in the literature~\cite{NLOpt}, it is
possible to improve the accuracy of results presented here by performing
matching at the NLO. However, our main goal here is to (re-)examine the influence of
the \np function on the $p_T$ distribution and for that purpose we consider
the LO matching sufficient. The NNLL resummed $b$ space resulted matched to
the NLO fixed-order expression was recently obtained in~\cite{BCDFG}.  
The problem of the negative large $p_T$ behaviour is normally solved by simply switching from the matched resummed
cross section to the fixed order cross section (at an appropriate value of
$p_T$) to obtain the cross section valid at all $p_T$.

At low values of $p_T$, the $b$ space predictions for the two choices of $g$ values are close
to each other, in  agreement with the conclusions of Ref.~\cite{BCS,BQ}.
The effect of changing the \np part is felt, albeit slightly, at all values of $\pt$. 
In contrast, all the predictions from the $p_T$ space formalism are already {\em equal}
at  $p_T \sim 15$~GeV. Around this value, the $p_T$ space formalism begins to return a `pure'
perturbative part (the Sudakov factor) without any \np contamination. 
It is clear from Eq.~\ref{KS:FNP}, that for $p_T \gg \ptlim$ and in the limit 
$\tilde{a} \rightarrow \infty$ the $p_T$ space expression for the resummed cross section 
has only purely perturbative content.
In the region of $\pt$ where the $\pt$ space predictions differ, the dependence on the values of 
the effective \np parameter $\tilde a$ is,
however, much stronger than the analogous dependence on the parameter $g$ in $b$ space.
For lower values of $\tilde a$ than the central value $\tilde a =0.06$ GeV$^{-2}$, the position 
of the peak and the value of the distribution at the peak changes significantly.  
In the case of the lower value of $\tilde a=0.032$  GeV$^{-2}$ the
value of the distribution at the peak is around 5\% larger than the
corresponding value for $\tilde a=0.06$ GeV$^{-2}$ and the peak location moves by about 4~GeV 
towards smaller values of $p_T$. For values of $\tilde a$ larger than the central value, 
the difference in the shape of the distributions is not so significant at moderate $p_T$.
\footnote{As an artefact of the method of implementing the \np contribution in the 
$\pt$ space formalism, the distribution becomes unphysical for large values of 
$\tilde a$ (corresponding to a strongly peaked distribution) at very small $p_T$.}
  
We believe that the difference between the $b$--space and $p_T$--space predictions 
for the central values of the \np coefficients in the small $p_T$ region originates in the
 essential differences in the methods for  accounting for \np effects in the two formalisms. 
In particular, the $b$ space formalism requires a prescription to regularize the argument of 
$\as$ at large $b$. This is achieved by introducing the quantity $b_*$ as in Eq.~\ref{bstar}. 
The $p_T$ space formalism, in turn, does not require any regularization prescription. 
The error band on the predictions of the Higgs $\pt$ distribution in the $\pt$ formalism, 
obtained by considering distributions for the two `extreme' values of $\tilde a$, 
i.e.  $\tilde a=0.032$ GeV$^{-2}$ and $\tilde a=0.308$ GeV$^{-2}$, is much larger than the 
corresponding error band from the $b$ space method. 
In particular, the lower value of  $\tilde a$ denotes the broadest \np gaussian 
component allowed, so that the contribution from the perturbative part is minimised 
in comparison with the other (allowed) values of $\tilde a$.   
However, it needs to be stressed that apart from intrinsic differences in the 
formalisms the methods of determining $g$ and $\tilde a$ coefficients for the Drell-Yan processes 
differ --- to obtain the effective $\tilde a$ coefficients we varied $\ptlim$ 
which resulted in relatively large errors on the values of the determined parameters. 
It is possible that a more accurate determination 
of the parameters would reduce the size of the error band on the predictions for the Higgs $\pt$ distribution 
in the $\pt$ space formalism. It is also true that allowing $b_{\rm lim}$ to vary while determining the
$g$ parameters would lead to a larger error band on the predictions from the $b$ space formalism. 
Nevertheless, the larger error band on the $\pt$ space results may suggest that the corresponding 
theoretical uncertainty
on the predictions obtained with the help of the $b$ space formalism is underestimated.


An alternative method for determining the values of the parameters for the $p_T$
space \np ansatz~(\ref{afnp}) relies on fixing $\ptlim$ while fitting data
for the effective parameters $\tilde a$. This method of finding the \np
coefficients for the $p_T$ space parameterization corresponds in a much more direct way
to the $b$ space analysis and  results in a significant decrease in the
errors bars on the value of the coefficients. However, when applied  it returns 
a considerably larger value of $\chi^2/d.o.f$ for the fits for the $\tilde
a_1,\;\tilde a_2$ and $\tilde a_3 $  parameters.
  
\section{Conclusions}

We have performed detailed fits of the resummed theoretical predictions to
the Drell-Yan data on the $p_T$ distributions for both the low $Q$ fixed target experiments and
 the $Z$ boson distribution data from the Tevatron $p \bar p$  collider. From these fits we established 
values of the \np components present 
in the resummed expressions. The fits were performed for two resummation formalisms: 
the well-known $b$ space formalism  and the $\pt$ space formalism. In both cases the \np parametrization 
is assumed to take the  form of a simple gaussian with a logarithmic dependence on $Q$ and $\sqrt s$.

The resummed expressions for the Higgs $\pt$ distribution in the gluon-gluon fusion production process also 
require a parametrization of \np effects. However, one may expect that the $gg$ initial state relevant for
Higgs production
will have  a different \np function supplementing the 
resummed expressions compared to the $q\bar q$ Drell-Yan process. We used a combination of theoretical 
intuition and a fit of the resummed distribution to data on $\Upsilon$ production in hadron-hadron collisions
  (which is also predominantly characterized by the $gg$ initial state) to determine the values of the
coefficients in the corresponding \np functions. To summarize, we see a difference in the peak region of the Higgs
transverse momentum distribution
between the $b$ and $\pt$ space predictions. In the latter case, the \np dependence is significant at 
lower $\pt$ but dies off rather rapidly with $\pt$. For the $b$ space prediction, in contrast, 
the \np variation is smaller at low $\pt$ but more persistent at higher $\pt$.

We believe that we are  unable to rule out
any of the predictions described in this study on the basis of theory and experiment, and therefore that
the corresponding span in the predictions is a measure of the true uncertainty on the theoretical
prediction for the small $p_T$ Higgs distribution. 
We note also that the uncertainty in the Higgs $\pt$ distribution due to the \np component in the 
$\pt$ space method seems to be larger than in the $b$ space method.

\vspace{1cm}

\noindent
{\bf Acknowledgments}\\
A.K. wishes to thank G.~Sterman and W.~Vogelsang for many discussions. 
A.K.'s research is supported by the U.S. Department of Energy (contract
no. DE-AC02-98CH10886).


\end{document}